\documentclass{llncs}

\usepackage{makeidx}
\usepackage{url}
\usepackage{mathfont}
\def\log{\mathop{{\rm log}}}

\input epsf

\newcommand{\OCfigure}[3]{
\begin{centering}
\begin{figure}
 \centerline{\vbox to #3 {\hfil
  \epsfysize=#3 \epsfbox{#1.eps}
  }}
 \caption{#2}
 \label{#1}
\end{figure}
\end{centering}}

\mathcode`O="724F

 \makeatletter
\def\hb@xt@{\hbox to }
\makeatother

\let\oldendproof\endproof
\def\endproof{\qed\oldendproof}

\makeatletter
\def\@begintheorem#1#2{\it \trivlist \item[\hskip \labelsep{\bf #1\ #2.\
}]}
\def\@opargbegintheorem#1#2#3{\it \trivlist
      \item[\hskip \labelsep{\bf #1\ #2\ (#3).}]}
\makeatother

\DeclareSymbolFont{AMSb}{U}{msb}{m}{n}
\DeclareSymbolFontAlphabet{\Bbb}{AMSb}
\def\R{\ensuremath{\Bbb R}}

\let\Real\R

\begin{document}

\addtocmark{Optimization Over Zonotopes and Training 
Support Vector Machines}

\mainmatter

\title{Optimization Over Zonotopes and \\
Training Support Vector Machines}

\author{Marshall Bern\inst{1} \and  David Eppstein\inst{2}}

\institute{Xerox PARC, 3333 Coyote Hill Rd., Palo Alto, CA 94304 \\
\email{bern@parc.xerox.com}
\and
Univ. of California, Irvine, Dept. Inf. \&
Comp. Sci., Irvine, CA 92697. \\
\email{eppstein@ics.uci.edu} }

\maketitle

\begin{abstract}
We make a connection between classical polytopes called
zonotopes and Support Vector Machine (SVM) classifiers.
We combine this connection with the ellipsoid method
to give some new theoretical results on training SVMs.
We also describe some special properties of $C$-SVMs
for $C \rightarrow \infty$. 
\end{abstract}

\section{Introduction}

A statistical classifier algorithm maps a set of 
{\it training vectors\/}---positively and negatively labeled points in 
$\Real^d$---to a decision boundary.
A Support Vector Machine (SVM) is a classifier algorithm
in which the decision boundary depends on only a subset
of training vectors, called the {\it support vectors\/}~\cite{Vapnik95}. 
This limited dependence on the training set
helps give SVMs good {\it generalizability\/},
meaning that SVMs are resistant to overtraining
even in the case of large $d$.
Another key idea associated with SVMs is the use 
of a {\it kernel function\/} in computing the dot product
of two training vectors. 
For example, the usual dot product $v \cdot w$
could be replaced by $k(v,w) = (v \cdot w)^2$ (quadratic kernel)
or by $k(v,w) = \exp(-\Vert v-w \Vert^2 )$ (radial basis function).
The kernel function~\cite{Scholkopf&99}
in effect maps the original training
vectors in $\Real^d$ into a higher-dimensional (perhaps infinite-dimensional)
{\it feature space\/} $\Real^{d'}$;
a linear decision boundary in $\Real^{d'}$ 
then determines a nonlinear decision surface back in $\Real^d$. 
For good introductions to SVMs see 
the tutorial by Burges~\cite{Burges98} or the book
by Cristianini and Shawe-Taylor~\cite{Cristianini&00}.

The basic {\it maximum margin\/} SVM applies to the case of
linearly separable training vectors, and divides
positive and negative vectors by a farthest-apart pair
of parallel hyperplanes, as shown in Figure~\ref{maxmargin}(a).
The decision boundary itself is typically the hyperplane
halfway between the  boundaries.
Computational geometers might expect that 
the extension of the SVM to the non-separable case would
divide positive and negative vectors by a least-overlapping
pair of half-spaces bounded by parallel hyperplanes,
as shown in Figure~\ref{maxmargin}(b). 
This generalization, however, may be overly sensitive to outliers, 
and hence the method of choice is a 
more robust {\it soft margin\/} classifier, called a $C$-SVM~\cite{Cortes&95,Vapnik95}
or $\nu$-SVM~\cite{Scholkopf&98} depending upon the precise formulation.
Parameter $C$ is a user-chosen penalty for errors. 

Computing the maximum margin classifier for $n$ vectors
in $\Real^d$ amounts to solving a quadratic program
(QP) with about $d$ variables and $n$ linear constraints. 
If the feature vectors are not {\it explicit\/} 
(that is, kernel functions are being used), 
then the usual Lagrangian formulation gives 
a QP with about $n+d$ variables and linear constraints.
Similarly, the soft margin classifier---with or without explicit
feature vectors---is computed in a Lagrangian formulation
with about $n+d$ variables and linear constraints.
The jump from $d$ to $n+d$ variables can have a great
impact on the running time and choice of QP algorithm.
Recent results in computational geometry~\cite{Gartner95,Matousek&92} give fast
QP algorithms for the case of large $n$ and small $d$, 
algorithms requiring about $O(nd) +  (\log n)\exp(O(\sqrt d \,))$ 
arithmetic operations.
The best bound on the number of arithmetic operations for a QP 
with $n+d$ variables and constraints is about
$O((n+d)^3 L)$, where $L$ is the precision 
of the input data~\cite{Todd97}. 

In this paper, we show that the jump from $d$ to $n+d$
is not necessary for soft margin classifiers with explicit
feature vectors.
More specifically, we describe training algorithms 
with running time near linear in $n$ and polynomial in $d$
and input precision, for two different scenarios: 
$C$ set by the user and $C \rightarrow \infty$.
The second scenario also introduces a natural measure
of separability of point sets.
Our algorithms build upon a geometric view
of soft margin classifiers~\cite{Bennett&00,Crisp&99} 
and the ellipsoid method for convex optimization.
Due to their reliance on explicit feature vectors 
and the ellipsoid method, and also due to the fact
that SVMs are more suited to the case of moderate $n$ 
and large $d$ than to the case of large $n$ and small $d$,
our algorithms have little practical importance. 
On the other hand, our results should be interesting theoretically.
We view the soft margin classifier as a problem defined over
a zonotope, a type of polytope that admits
an especially compact description.
Accordingly, our algorithms have lower complexity than either
the vertex or facet descriptions of the polytopes.

\OCfigure{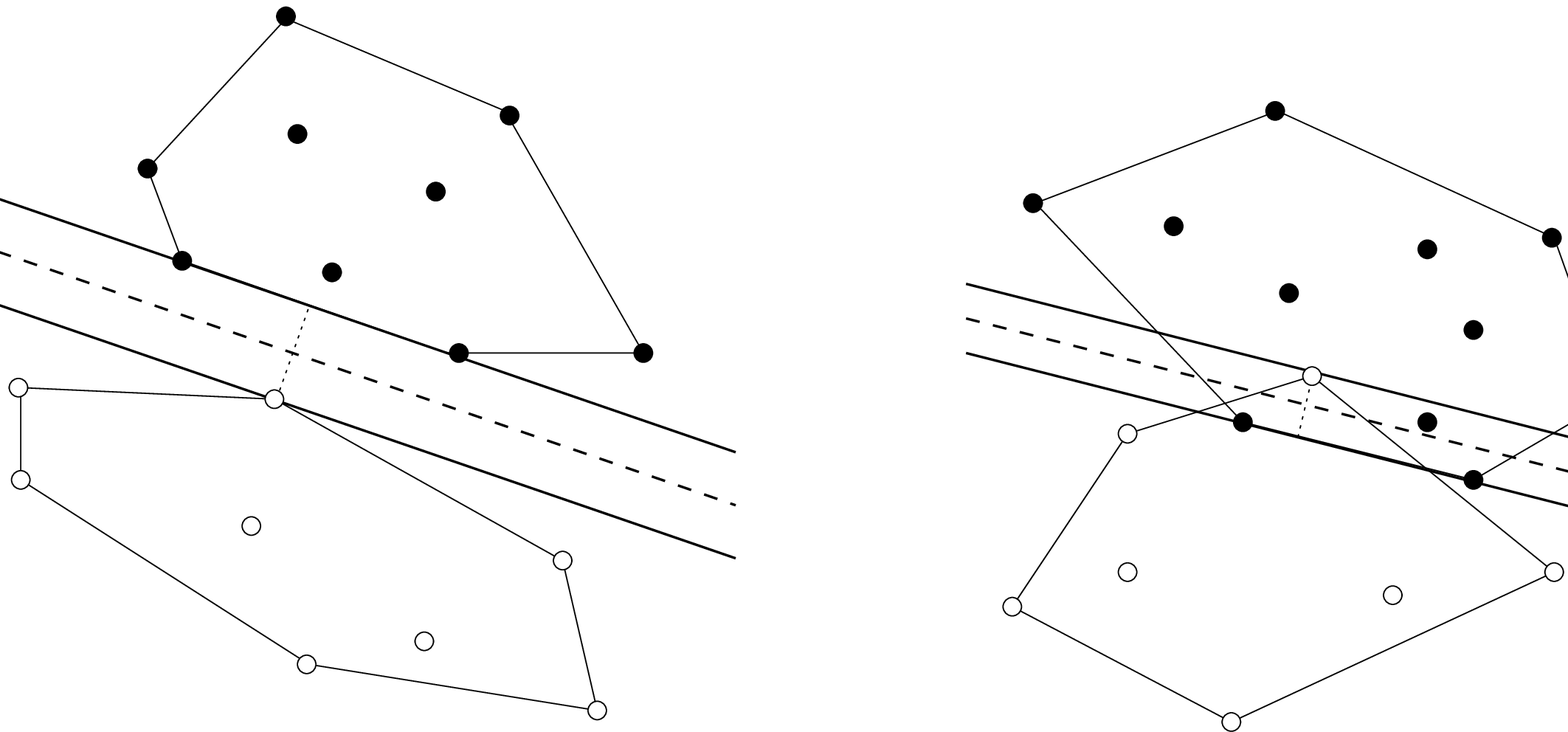}
{(a) The maximum margin SVM classifier for the separable case.
The dashed line shows the decision boundary.
(b) The most natural generalization to the non-separable case
is not popular.}{2.0in}

\section{SVM Formulations}\label{nuform}

We adopt the usual SVM notation
and mostly follow the presentation
of Bennett and Bredensteiner~\cite{Bennett&00}. 
The training vectors are $x_1, x_2, \dots , x_n$, points in $\Real^d$.
The corresponding labels are $y_1, y_2, \dots , y_n$, each of which is
either $+1$ or $-1$.
Let $I_+ = \{ \, i ~|~ y_i = +1 \, \}$
and $I_- = \{ \, i ~|~ y_i = -1 \, \}$.
We use $w$ and $x$ to denote vectors in $\Real^d$ and
$b$ to denote a scalar.  
We use the dot product notation $w \cdot x$, but in this section
$w\cdot x$ could be standing in for the kernel function $k(w,x)$.

In the maximum margin SVM we seek parallel hyperplanes
defined by the equations $w \cdot x = b_+$ and
$w \cdot x = b_-$ such that 
$w \cdot x_i \leq b_-$ for all $i \in I_-$
and 
$w \cdot x_i \geq b_+$ for all $i \in I_+$.
The signed distance between these two hyperplanes---the {\it margin\/}---is
${b_+ - b_-} \over {\Vert w \Vert }$ and hence
can be maximized by minimizing $\Vert w \Vert^2 - (b_+ - b_-)$.
\begin{eqnarray}\label{maxmarg}
\min_{w, \, b_+, \, b_-} ~~  \Vert w \Vert^2 - (b_+ - b_-) & {\rm ~subject~to } \\ \nonumber
x_i \cdot w \geq  b_+ ~~{\rm  for }~~  i \in I_+,  & \qquad
x_i \cdot w \leq b_-  ~~{\rm  for }~~  i \in I_-.
\end{eqnarray}
A popular choice for the decision boundary is the plane
halfway between the parallel hyperplanes,
$w \cdot x = (b_+ + b_-)/2$, and hence
each unknown vector $x$ is classified according to the sign of 
$w \cdot x - (b_+ + b_-) /2$.

In the linearly separable case, we can
set $b_+ = 1-b$ and $b_- = -1-b$ (thereby rescaling $w$) and 
obtain the following optimization problem, the
standard form in most SVM treatments~\cite{Burges98}.
\begin{eqnarray}\label{hardsvm}
\min_{w, \, b} ~~  \Vert w \Vert^2 & {\rm ~subject~to } \\ \nonumber
x_i \cdot w + b  \geq  1  ~~{\rm  for }~~  i \in I_+, & \quad\qquad
x_i \cdot w + b \leq  -1  ~~{\rm  for }~~  i \in I_-.
\end{eqnarray}
Notice that this QP has  $d+1$ variables and $n$ linear constraints. 
At the solution, $w$ is a linear combination of $x_i$'s,
$2/ \Vert w \Vert$ gives the margin, 
and $w \cdot x + b = 0$ gives the halfway decision boundary.

The dual problem to maximizing the distance between
parallel hyperplanes separating the positive and
negative convex hulls is to minimize the distance 
between points inside the convex hulls.
Thus the dual in the separable case
is the following.
\begin{equation}\label{minseg}
\min_{\alpha_i}\, 
\Bigl\Vert \sum_{i \in I_+} \alpha_i x_i - \sum_{i \in I_-} \alpha_i x_i 
\Bigr\Vert ^2
~ {\rm subject~to} ~~
0 \leq \alpha_i \leq 1, 
~ \sum_{i \in I_+} \alpha_i = 1,
~ \sum_{i \in I_-} \alpha_i = 1.
\end{equation}
Karush-Kuhn-Tucker (complementary slackness) conditions
show that the optimizing value of $w$ 
for~(\ref{maxmarg}) is given by the optimizing values
 of $\alpha_i$ for~(\ref{minseg}):
$w =  \sum_{i\in I_+} \alpha_i x_i - \sum_{i\in I_-} \alpha_i x_i$. 
The vectors $x_i$ with $\alpha_i > 0$ are called the {\it support vectors\/}.

The soft margin SVM adds slack variables to formulation~(\ref{maxmarg}),
and then penalizes solutions proportional to the sum of these variables.
Slack variable $\xi_i$ measures the {\it error\/} for training vector $x_i$,
that is, how far $x_i$ lies on the wrong side of the parallel hyperplane
for $x_i$'s class.
\begin{eqnarray}\label{softmarg}
\min_{w, \, b_+, \, b_-, \, \xi_i} ~~   \Vert w \Vert^2 + (b_+ - b_-) +
 \mu\sum_{i=0}^n \xi_i &
 {\rm ~subject~to } \\ \nonumber
\xi_i \geq 0 ~~\forall i,
\quad x_i \cdot w  \geq b_+  - \xi_i  ~\,{\rm  for }~\,  i \in I_+, & 
~~ 
x_i \cdot w  \leq  b_- + \xi_i  ~\,{\rm  for }~\,  i \in I_-.
\end{eqnarray}
The standard $C$-SVM formulation~\cite{Vapnik95} 
again sets $b_+ = 1-b$ and $b_- = -1-b$.
\begin{eqnarray}\label{csvm}
\min_{w, \, b, \, \xi_i} ~~    \Vert w \Vert^2 + C\sum_{i=0}^n \xi_i &
 {\rm ~subject~to } \\ \nonumber
\xi_i \geq 0 ~~\forall i, ~~
x_i \cdot w + b \geq 1 - \xi_i  ~\,{\rm  for }~\,  i \in I_+, & ~~
x_i \cdot w + b \leq  -1 + \xi_i  ~\,{\rm  for }~\,  i \in I_-.
\end{eqnarray}

In formulation~\ref{csvm}, the decision boundary is $w \cdot x = b$.
Formulation~(\ref{softmarg}), however,
does not set the decision boundary, but only its direction. 
Crisp and Burges~\cite{Crisp&99}
write that because ``originally the sum of $\xi_i$'s term
arose in an attempt to approximate the number of errors'',
the best option might be to run a ``simple line search''
to find the decision boundary that actually minimizes
the number of training set errors. 

The dual of formulation~(\ref{softmarg})
in the separable case minimizes the distance between points inside 
``reduced'' or ``soft'' convex hulls~\cite{Bennett&00,Crisp&99}.
\begin{equation}\label{minsoft}
\min_{\alpha_i}\, 
\Bigl\Vert \sum_{i \in I_+} \alpha_i x_i - \sum_{i \in I_-} \alpha_i x_i
\Bigr\Vert ^2
~{\rm subject~to}~~
0 \leq \alpha_i \leq \mu, ~ \sum_{i\in I_+} \alpha_i = 1, 
~ \sum_{i\in I_-} \alpha_i = 1.
\end{equation}
See Figure~\ref{non-sep}. 
The {\it reduced convex hull\/} of points $x_i$, $i \in I_+$,
is the set of convex combinations
of $\alpha_i x_i$ with each $\alpha_i \leq \mu$. 
(Notice that in~(\ref{softmarg}) there is no reason to consider $\mu > 1$.) 
We shall say more about reduced convex hulls
in the next section.

\OCfigure{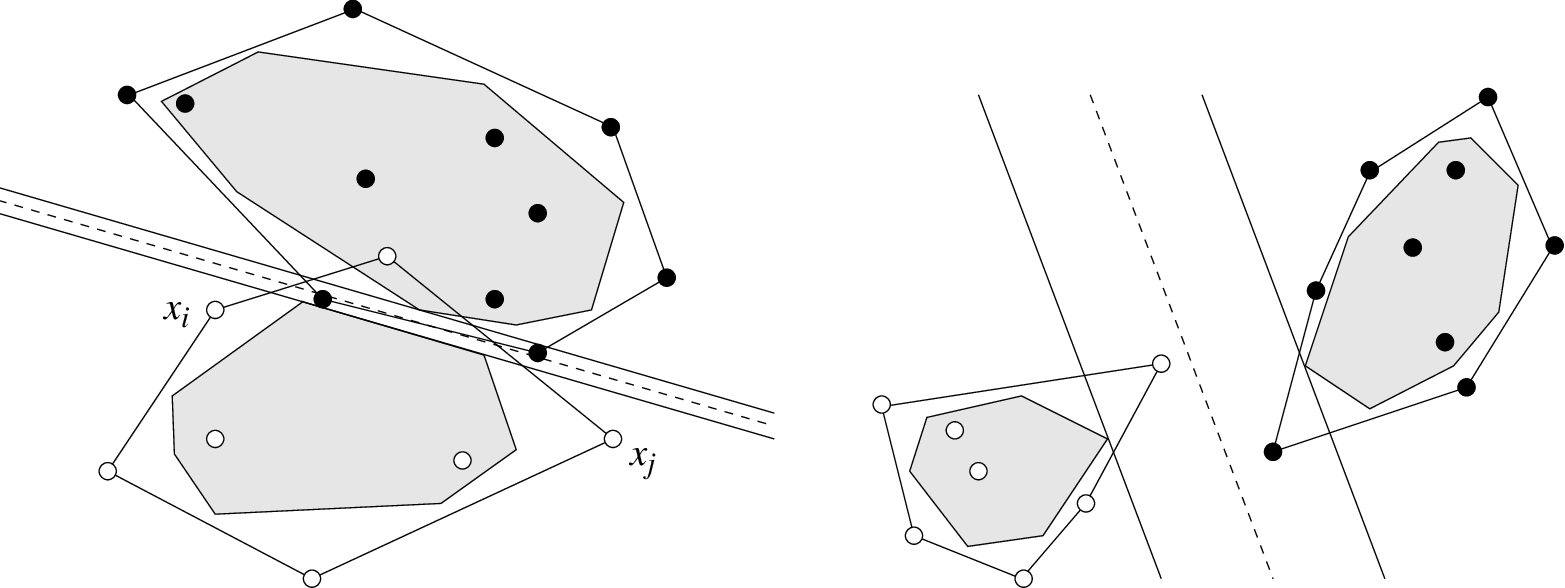}
{(a) Soft margin SVMs maximize the margin between
reduced convex hulls.  
(b) Although the soft margin is often explained
as a way to handle non-separability, it can help
in the separable case as well.}
{1.8in}

The dual view highlights a slight difference between
formulations~(\ref{softmarg}) and~(\ref{csvm}).
Formulation~(\ref{softmarg}) allows the direct setting of the
reduced convex hulls.  Parameter $\mu$ limits the influence
of any single training point; if the user expects no more than 
four outliers in the training set, then an appropriate choice
of $\mu$ might be $1/9$ in order to ensure that the
majority of the support vectors are non-outliers.
If the reduced convex hulls intersect, 
the solution to~(\ref{softmarg}) is
the least-overlapping pair of half-spaces, as in Figure~\ref{maxmargin}(b).
Formulation~(\ref{csvm}) is also always feasible---unlike
the standard hard margin formulation~(\ref{hardsvm})---but 
it never allows the reduced convex hulls to intersect.
As $C \rightarrow \infty$ the reduced convex hulls 
either fill out their convex hulls (the separable case)
or continue growing until they asymptotically
touch (the non-separable case).

\section{Reduced Convex Hulls and Zonotopes }

Assume $0 \leq \mu \leq 1$ 
and define the positive and negative reduced convex hulls by 
$$
H_{+\mu} = \Bigl\{ \sum_{i \in I_+} \alpha_i x_i ~\Big| 
\sum_{i \in I_+} \alpha_i = 1, ~~ 0 \leq \alpha_i \leq\mu \Bigr\},
$$
$$
H_{-\mu} = \Bigl\{ \sum_{i \in I_-} \alpha_i x_i ~\Big| 
\sum_{i \in I_-} \alpha_i = 1, ~~ 0 \leq \alpha_i \leq\mu \Bigr\} .
$$
Figure~\ref{softhull} shows the reduced convex hull
of three points $x_1$, $x_2$, and $x_3$ for various values of $\mu$.
The reduced convex hull grows from the centroid at $\mu = 1/3$
to the convex hull at $\mu = 1$; for $\mu < 1/3$ it is empty.
In Figure~\ref{non-sep}, $\mu$ is a little less than $1/2$.

A reduced convex hull is a special case of 
a {\it centroid polytope\/}, the locus of
possible weighted averages of points each with
an unknown weight within a certain range~\cite{Bern&95}. 
For reduced convex hulls, each weight $\alpha_i$ has the same range $[0, \mu]$
and the sum of the weights is constrained to be $1$.
In~\cite{Bern&95} we related centroid polytopes in 
$\Real^d$ to special polytopes, called zonotopes, in $\Real^{d+1}$.
We repeat the connection here, specialized to the case
of reduced convex hulls.

\OCfigure{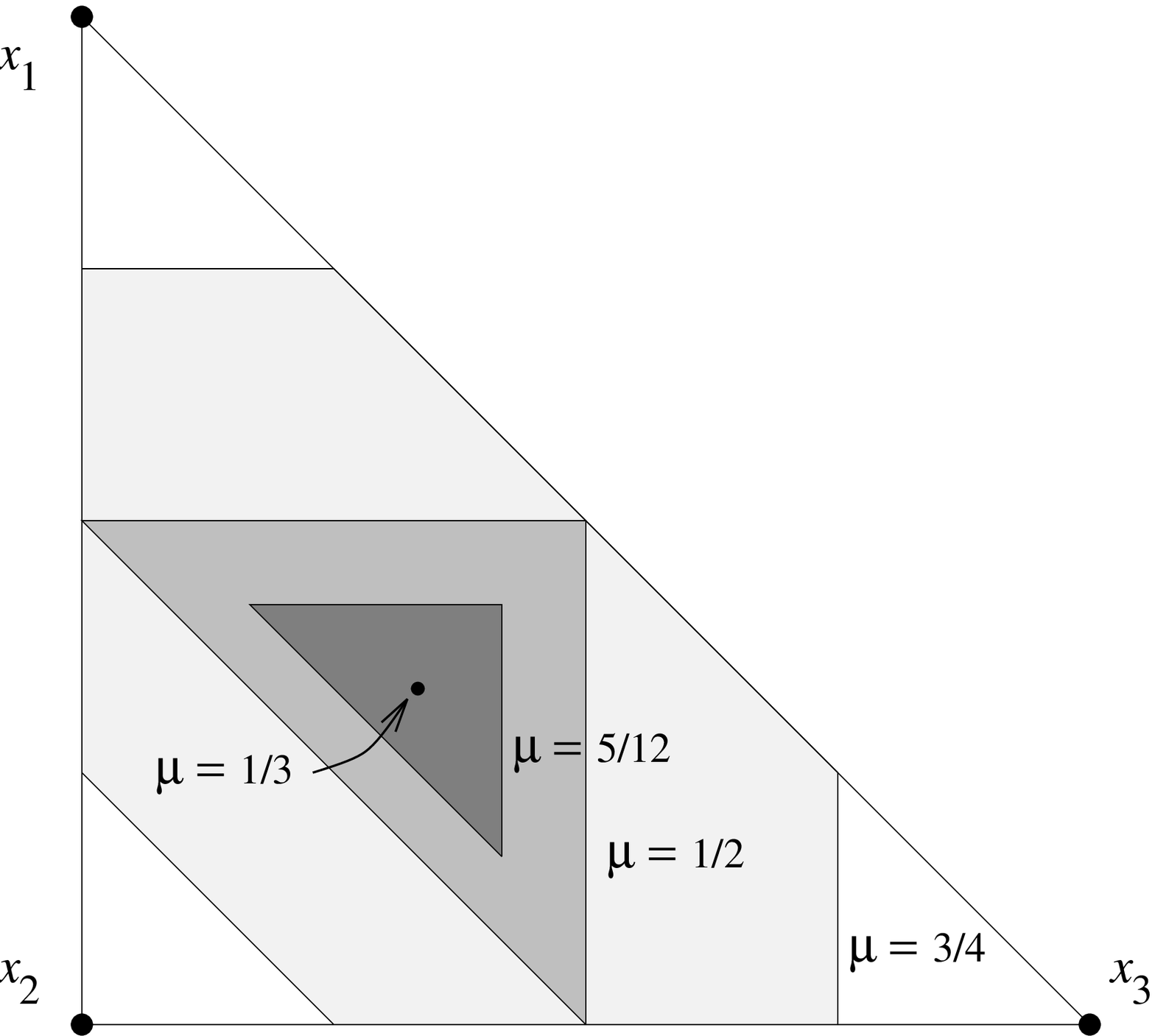}
{The reduced convex hull of 3 points ranges from the centroid 
to the convex hull.}
{2.0in}

Let $v_i$ denote $(x_i, 1)$, the vector in $\Real^{d+1}$ that
agrees with $x_i$ on its first $d$ coordinates
and has 1 as its last coordinate.  
Define
$$
Z_{+\mu} =
 \Bigl\{\,\sum_{i \in I_+} \alpha_i v_i  ~\Big|~ 0 \leq \alpha_i \leq \mu \,\Bigr\}.
$$
Polytope $Z_{+\mu}$ is a 
Minkowski sum\footnote{
The {\it Minkowski sum\/} of  sets $A$ and $B$ in $\R^{d+1}$
is $\{ p + q ~|~ p \in A {\rm ~and~} q \in B \}$.
}
of line segments of the form 
$S_i = \{ \, \alpha_i v_i ~|~ 0 \leq \alpha_i \leq \mu \,\}$.
The Minkowski sum of line segments is a
special type of convex polytope called a 
{\it zonotope\/}~\cite{Bern&95,Edelsbrunner87}.
Polytope $H_{+\mu}$ is the cross-section of $Z_{+\mu}$
with the $(d+1)$-st coordinate (which by construction is also 
$\sum_i \alpha_i$) equal to one. 
Of course, $H_{-\mu}$ can also be related to a zonotope in the same way.
The following lemmas state the property
of zonotopes and reduced convex hulls that underlies our algorithms.
Lemma~\ref{softopt} is implicit in 
Keerthi et al.'s iterative nearest-point approach to 
SVM training~\cite{Keerthi&99}.

\begin{lemma}\label{zonopt}
Let $Z$ be a zonotope that is the Minkowski
sum of $n$ line segments in $\Real^d$.
There is an algorithm with $O(nd)$ arithmetic operations
for optimizing a linear function over $Z$.
\end{lemma}

\begin{proof}
Assume that we are trying to find a vertex $v$ in zonotope $Z$
extreme in direction $w$, that is,
that maximizes the dot product $w \cdot v$.
Assume that $Z$ is the Minkowski sum of line segments of the form 
$S_i = \{ \, \alpha_i v_i ~|~ 0 \leq \alpha_i \leq \mu \,\}$, 
where $v_i \in \Real^{d+1}$.
We simply set each $\alpha_i$ independently to $0$ or $\mu$,
depending upon whether the projection
of $v_i$ onto $w$  is negative or positive. 
\end{proof}

\begin{lemma}\label{softopt}
There is an algorithm with $O(nd)$ arithmetic operations
for optimizing a linear function over a reduced convex hull
of $n$ points in $\Real^d$.
\end{lemma}

\begin{proof}
Assume that we are trying to find a vertex $x$ in zonotope $H_{+\mu}$
extreme in direction $w$.
Order the $x_i$'s with $y_i = +1$ according
to their projection onto vector $w$, breaking ties arbitrarily. 
In decreasing order by projection along $w$, 
set the corresponding $\alpha_i$'s to $\mu$
until doing so would violate the constraint
that $\sum_{i\in I_+} \alpha_i = 1$.
Set $\alpha_i$ for this ``transitional'' vector
to the maximum value allowed by this constraint, and
finally set the remaining $\alpha_i$'s to $0$.
Then $x = \sum_{i\in I_+} \alpha_i x_i$
maximizes $w \cdot x$.
\end{proof}

An interesting combinatorial question asks for the worst-case
complexity of a reduced convex hull $H_{+\mu}$. 
The vertex $x$ of $H_{+\mu}$ that is extreme for direction $w$
can be associated with the set of $x_i$'s for which $\alpha_i > 0$.
If $\mu = 1/k$, then as in Lemma~\ref{softopt},
$x$'s set is the first $k$ points in direction $w$,
a set of $k$ points that can be separated from the other $n-k$ points by a 
hyperplane normal to $w$.
And conversely, each separable set of $k$ points defines a unique vertex of 
$H_{\mu}$.  Hence the maximum number of vertices of $H_{+\mu}$ is  
equal to the maximum number of {\it $k$-sets\/} for $n$ points
in $\Real^d$, which is known to be 
$\omega(n^{d-1})$ and $o(n^d)$~\cite{Toth00,Zivaljevic&92}. 
In~\cite{Bern&95} we showed that a more general centroid polytope 
in which each point $x_i$ has $\alpha_i$ between $0$ and $\mu_i$
(that is, different weight bounds for different points)
may have complexity $\Theta(n^d)$.

We can also apply the argument in the proof of Lemma~\ref{softopt} 
to say something about the optimizing values of the variables 
in~(\ref{softmarg}) and~(\ref{minsoft}) for the non-separable case. 
(Alternatively we can derive the same statements 
from the Karush-Kuhn-Tucker conditions.) 
Each of $H_+$ and $H_-$ has a transition in
the sorted order of the $x_i$'s when projected along the normal $w$
to the parallel pair of hyperplanes.
For $x_i$ with $i \in I_+$, $\alpha_i = 0$ if $x_i \cdot w$
lies on the ``right'' side of the transition,
$0 \leq \alpha_i \leq \mu$ if $x_i \cdot w$ coincides with
the transition, and 
$\alpha_i = \mu$ if $x_i$ lies on the ``wrong'' side of the transition.
Of course an analogous statement holds for $x_i$ for $i \in I_-$. 
As usual, the support vectors are those $x_i$ with $\alpha_i > 0$. 
Thus all training set errors are support vectors. 
In Figure~\ref{non-sep}(a) there are six support vectors:
two transitional unfilled dots (marked $x_i$ and $x_j$)
and one wrong-side unfilled dot, 
along with one transitional and two wrong-side filled dots.

\section{Ellipsoid-Based Algorithms}

We first assume that $\mu$ has been fixed in advance, 
perhaps using some knowledge of the expected number of outliers
or the desired number of support vectors. 
We give an algorithm for solving formulation~(\ref{minsoft}).

One approach would be to compute the vertices of
$H_{+\mu}$ and $H_{-\mu}$ and then use formulation~(\ref{maxmarg})
with positive and negative training vectors replaced by the vertices of 
$H_{+\mu}$ and $H_{-\mu}$ respectively.
However, the number of vertices of 
$H_{+\mu}$ and $H_{-\mu}$ may be very large, so this 
algorithm could be very slow.

So instead we exploit a polynomial-time equivalence
between separation and optimization 
(see for example~\cite{Schrijver}, chapter 14.2).
The input to the {\it separation problem\/} is a point
$q$ and a polytope $P$ (typically given by a system of linear inequalities). 
The output is either a statement
that $q$ is inside $P$ or a hyperplane separating $q$ and $P$.
The input to the {\it optimization problem\/} is a direction $w$ 
and a polytope $P$. 
The output is either a statement that $P$ is empty,
a statement that $P$ is unbounded in direction $w$,
or a point in $P$ extreme for direction $w$. 
The two problems are related by projective duality,\footnote{
The more famous direction of this equivalence is
that separation---which can be solved directly by 
checking each inequality---implies optimization. 
This result is a corollary of Khachiyan's ellipsoid method.
}
and a subroutine for solving one can be used to solve the other
in a number of calls that is polynomial in the dimension $d$ 
and the input precision, that is, the number of bits
in $q$ or $w$ plus the  maximum number of
bits in an inequality defining $P$.

In our case, the polytope is not given by inequalities, but
rather as a Minkowski sum of line segments; this presentation
has an impact on the required precision. 
If the input precision is $L$, the maximum number of bits
in one of the feature vectors $x_i$, then the
maximum number of bits in a vertex of the polytope
is $O(d^2 L \log n)$. 
What is new is the $O(\log n)$ term, resulting
from the fact that a vertex of a zonotope is a sum
of up to $n$ input vectors.

\begin{theorem}\label{fixedmu}
Given $n$ explicit feature vectors in $\Real^d$ and 
$\mu$ with $1/n \leq \mu \leq 1$, 
there is a polynomial-time algorithm for computing
a soft margin classifier, with the number of arithmetic
operations linear in $n$ and polynomial in $d$, $L$, and $\log n$.
\end{theorem}

\begin{proof}
As in~\cite{Keerthi&99}, consider the polytope $P$
that is the Minkowski sum of 
$H_{+\mu}$ and $-H_{-\mu}$, that is, 
$P = \{ \, v_+ - v_- ~|~ 
v_+ \in H_{+\mu} ~{\rm and}~  v_- \in H_{-\mu} \, \}$.
We are trying to minimize over $P$ the convex quadratic 
objective function $ \Vert v  \Vert^2$, that
is, the length of a line segment between $H_{+\mu}$ and $H_{-\mu}$.

For a given direction $w$, we can find the solution 
$v = v_+ - v_-$ to the linear optimization problem for $P$
by using Lemma~\ref{softopt} to find the 
$v_+$ optimizing $w$ over $H_{+\mu}$ 
and the $v_-$ optimizing $w$ over $H_{-\mu}$.
Now given a point $q \in \Real^d$, we can use 
this observation and the polynomial-time equivalence
between separation and optimization
to solve the separation problem for $q$ and $P$ in time 
linear in $n$ and polynomial in $d$ and $L$.
We can use this solution to the separation problem for $P$
as a subroutine for the ellipsoid method
(see~\cite{Kozlov,Schrijver}) in order to optimize $ \Vert v  \Vert^2$ over $P$.
Given an optimizing choice of $v = v_+ - v_-$, it is easy
to find the best pair of parallel hyperplanes
and a decision boundary, either the $C$-SVM decision boundary
or some other reasonable choice within the parallel family. 
\end{proof}

Now assume that we are in the non-separable case.
We shall show how to solve for the maximum $\mu$
for which the reduced convex hulls have non-intersecting interior, 
that is, the $\mu$ for which the margin is $0$. 
This choice of $\mu$ corresponds to $C \rightarrow \infty$
and the objective function simplifying to $\sum_i \xi_i$ 
in formulation~(\ref{csvm}). 

This choice of $\mu$ has two special properties.
First, among all settings of $C$, $C \rightarrow \infty$
tends to give the fewest support vectors.
To see this, imagine shrinking the shaded regions
in Figure~\ref{non-sep}(a).
Support vectors are added each time one of the parallel hyperplanes
crosses a training vector. On the other hand,
a support vector may be lost occasionally when the number of 
reduced convex hull vertices on the parallel hyperplanes changes,
for example, if the vertex supporting the upper
parallel line in Figure~\ref{non-sep}(a) slipped off to the right
of the segment supporting the lower parallel line.

Second, the $\mu$ for which the margin is zero gives a natural
measure of the separability of two point sets. 
For simplicity, let $|I_+| = |I_-| = n/2$
and normalize the zero-margin $\mu$ by $\mu^* =  (\mu - 2/n)/(1-2/n)$.
The separability measure 
$\mu^*$ runs from 0 to 1, with 0 meaning that the centroids
coincide and 1 meaning that the convex hulls have disjoint interiors.
Computing the zero-margin $\mu$ as the
maximum value of a dual variable $\alpha_i$ using
formulation~(\ref{csvm}) above is no harder than training a $C$-SVM,
and in the case of explicit features, it should
be significantly easier, as we now show. 

We can formulate the problem as 
minimizing $\mu$ subject to 
$$
\sum_{i \in I_+} \beta_i x_i =
\sum_{i \in I_-} \beta_i x_i , \quad\qquad 
\sum_{i \in I_+} \beta_i = 1 , \quad\qquad
\sum_{i \in I_-} \beta_i = 1 , \quad\qquad
0 \leq \beta_i \leq \mu.
$$ 
As above, let $v_i$ denote $(x_i, 1)$, the vector in $\Real^{d+1}$ that
agrees with $x_i$ on its first $d$ coordinates
and has 1 as its last coordinate.  
Letting $\alpha_i = \beta_i/\mu$, we
can rewrite the problem as maximizing 
$$
1/\mu = \sum_{i \in I_+} \alpha_i =  \sum_{i \in I_-} \alpha_i
$$
subject to
\begin{equation}\label{mulin}
\sum_{i \in I_+} \alpha_i v_i = 
\sum_{i \in I_-} \alpha_i v_i , \quad\qquad 
0 \leq \alpha_i \leq 1.
\end{equation}
Yet another way to state the problem is to ask for
the point with maximum $(d+1)$-st coordinate
in $Z_+ \cap Z_-$, where
$$
Z_+ = \Bigl\{\,\sum_{i \in I_+} \alpha_i v_i  ~\Big|~ 0 \leq \alpha_i \leq 1 \,\Bigr\},
\qquad
Z_- = \Bigl\{\,\sum_{i \in I_-} \alpha_i v_i  ~\Big|~ 0 \leq \alpha_i \leq 1 \,\Bigr\}.
$$
Polytopes $Z_+$ and $Z_-$ are each zonotopes, Minkowski sums
of line segments of the form 
$S_i = \{ \, \alpha_i v_i ~|~ 0 \leq \alpha_i \leq 1 \,\}$.

\begin{theorem}
Let $Z_1$ and $Z_2$ be zonotopes defined by a total of 
$n$ line segments in $\Real^d$.
There is an algorithm for optimizing 
a linear objective function over $Z_1 \cap Z_2$,  
with the number of arithmetic operations linear in $n$
and polynomial in $d$, $L$, and $\log n$.
\end{theorem}

\begin{proof}
Given a point $q$ and zonotope $Z_i$, $1 \leq i \leq 2$, we can use
Lemma~\ref{zonopt} and the polynomial-time equivalence
between separation and optimization
to solve the separation problem for $q$ and $Z_i$ in time 
linear in $n$ and polynomial in $d$, $L$ and $\log n$.
We can solve the separation problem for the intersection
of zonotopes $Z_1 \cap Z_2$ simply by solving it separately for each zonotope. 
We now use the equivalence between separation and optimization
in the other direction 
to conclude that we can also solve the optimization problem
for an intersection of zonotopes. 
\end{proof}

The proof of the following result then follows
from the ellipsoid method in the same way as the proof
of Theorem~\ref{fixedmu}. 

\begin{corollary}\label{corol}
Given $n$ explicit feature vectors in $\Real^d$,
there is a polynomial-time algorithm for computing 
the maximum $\mu$ for which $H_{+\mu}$ and $H_{-\mu}$ are linearly separable,
with the number of arithmetic operations linear in $n$
and polynomial in $d$, $L$, and $\log n$.
\end{corollary}

Theorem~\ref{fixedmu} and Corollary~\ref{corol} can be extended to some 
cases of implicit feature vectors.  For example, the quadratic kernel 
$k(v,w) = (v \cdot w)^2$ for vectors $v = (v_1, v_2)$
and $w = (w_1, w_2)$ in $\Real^2$
is equivalent to an ordinary dot product in 
$\Real^3$, namely $k(v,w) = \Phi(v) \cdot \Phi(w)$, where
$\Phi(v) = \left( v_1^2, \sqrt{2} v_1v_2, v_2^2 \right)$.
In general~\cite{Burges98},
a polynomial kernel $k(v,w) = (v \cdot w)^p$  amounts to lifting the training
vectors from $\Real^d$ to $\Real^{d'}$ where $d' = {{d+p-1} \choose p}$.
Radial basis functions, however, give $d' = \infty$, and
the SVM training problem seems to necessarily involve $n+d$ variables.
(The rather amazing part is that it is a combinatorial
optimization problem at all!)

\section{Discussion and Conclusions}

In this paper we have connected SVMs to 
some recent results in computational geometry and mathematical
programming.  These connections raise some new 
questions, both practical and theoretical.

Currently the best practical algorithms for training SVMs, 
Platt's sequential minimal optimization (SMO)~\cite{Platt98} 
and Keerthi et al.'s  nearest point algorithm (NPA)~\cite{Keerthi&99},
can be viewed as interior-point methods that iteratively
optimize the margin over line segments. 
Both algorithms make use of heuristics to find line segments
close to the exterior, meaning line segments with
$\alpha_i$ weights set to either 0 or $C$.

Computational geometry may have a practical algorithm to contribute for
the case of $n$ large and $d$ small, 
say $n \approx 100,000$ and $d \approx 20$: 
the generalized linear programming (GLP) paradigm of
Matou\v{s}ek et al.~\cite{Gartner95,Matousek&92}.
The training vectors need not actually live in $\Real^d$
for small $d$, so long as the GLP dimension of the problem is small,
where the GLP dimension is the number of support vectors
in any subproblem defined by a subset of the training vectors.

On the theoretical side, we are wondering about the existence
of strongly polynomial algorithms for QP problems over zonotopes.  
Due to the combinatorial equivalence of zonotopes and arrangements, 
the graph diameter of a zonotope is known to be only $O(n)$; polynomial
graph diameter is of course a necessary condition
for the existence of a polynomial-time simplex-style algorithm. 

\section*{Acknowledgments}

David Eppstein's work was done in part while visiting Xerox PARC,
and supported in part by NSF grant CCR-9912338. 
We would also like to thank Yoram Gat for a number of helpful discussions.

\bibliographystyle{abuser}

\end{document}